\documentclass[pra,aps,twocolumn,showpacs,nofootinbib]{revtex4}

\usepackage{amsfonts,amsmath,bm,amsthm,amssymb}
\usepackage{bbold}
\usepackage[OT4]{fontenc}
\usepackage{color}
\usepackage{soul}

\usepackage{graphicx}
\usepackage{epstopdf}

\usepackage{hyperref}
\hypersetup{
   colorlinks=true,         
   linkcolor=blue,          
   citecolor=blue,          
   urlcolor=Violet          
}
\usepackage{bookmark}

\newcommand{\bra}[1]{\ensuremath{\left\langle#1\right|}}
\newcommand{\ket}[1]{\ensuremath{\left|#1\right\rangle}}
\newcommand{\braket}[2]{\ensuremath{\left\langle#1 \vphantom{#2}\middle|  #2 \vphantom{#1}\right\rangle}}
\newcommand{\ketbra}[2]{\ensuremath{\left|#1\right\rangle\!\left\langle#2\right|}}

\newcommand{\tr}[1]{\mathrm{Tr}\left( #1 \right)}
\newcommand{\iden}{\mathbb{1}}

\renewcommand{\v}[1]{\ensuremath{\underline{\boldsymbol #1}}}

\newtheorem{existence_theorem}{Theorem}
\newtheorem{ZNZD_existence}{Lemma}
\newtheorem{definition}{Definition}

\begin{document}

\title{Operational constraints on state-dependent formulations of quantum error-disturbance trade-off relations} 

\author{Kamil Korzekwa}
\author{David Jennings}
\author{Terry Rudolph}
\affiliation{Department of Physics, Imperial College London, London SW7 2AZ, United Kingdom}

\begin{abstract}

We argue for an operational requirement that all state-dependent measures of disturbance should satisfy. Motivated by this natural criterion, we prove that in any $d$-dimensional Hilbert space and for any pair of non-commuting operators, $A$ and $B$, there exists a set of at least $2^{d-1}$ zero-noise, zero-disturbance (ZNZD) states, for which the first observable can be measured without noise and the second will not be disturbed. Moreover, we show that it is possible to construct such ZNZD states for which the expectation value of the commutator $[A,B]$ does not vanish. Therefore any state-dependent error-disturbance relation, based on the expectation value of the commutator as a lower bound, must violate the operational requirement. We also discuss Ozawa's state-dependent error-disturbance relation in light of our results and show that the disturbance measure used in this relation exhibits unphysical properties. We conclude that the trade-off is inevitable only between state-independent measures of error and disturbance.

\end{abstract}


\maketitle

\section{Introduction}

Despite almost a century of research on quantum theory, one of its fundamental building blocks, the quantum measurement process, is still actively investigated. One of the earliest results in this field is the famous Heisenberg uncertainty relation. Its best known modern formulation (also known as the Heisenberg-Robertson uncertainty relation \cite{robertson1929uncertainty}) concerns the outcome statistics of two independent measurements of noncommuting observables performed on an ensemble of identically prepared quantum states. It states
that the product of variances of these two outcome statistics is lower-bounded by the mean value of the commutator of measured observables in the given quantum state. Although this formulation says nothing about the effect of one measurement on the outcome statistics of the other, it is often misinterpreted in the spirit of the original Heisenberg microscope thought experiment \cite{heisenberg1949physical}, i.e., that the bigger the precision of the measurement of one observable, the bigger the disturbance to a subsequent measurement of the other one (with which it does not commute). Indeed, the formulation of uncertainty relation in terms of precision and disturbance of sequential measurements was Heisenberg's original concept and is sometimes called the error-disturbance uncertainty relation. Interestingly, it was not until recently that the problem of sequential measurements \cite{ozawa2003universally,busch2013proof} and joint measurements \cite{ozawa2004uncertainty,hall2004prior} has been addressed with a mathematically rigorous approach and in recent months has become the topic of much discussion \cite{rozema2013note,ozawa2013disproving,dressel2013certainty,busch2013heisenberg, bastos2014robertson,ipsen2013error,busch2013noise,branciard2013error,ringbauer2013joint}. 

Most of the controversies around the error-disturbance relation arise due to disagreement about proper definitions of error (noise) and disturbance (a detailed review of most commonly used notions can be found in \cite{busch2004noise}). In this paper, instead of proposing new definitions, we try to clarify the subject of sequential quantum measurements in finite-dimensional Hilbert spaces by examining the consequences of the commonly accepted \cite{busch2004noise,busch2013proof} and operationally motivated requirement that all physically meaningful notions of disturbance should satisfy. Specifically, we focus on operationally detectable disturbances for which it is natural to define

\begin{definition} (Operational disturbance). Consider a nonselective measurement of observable $A$ on a system in state $\rho$ that results in final state $\rho '$. We say the measurement of $A$, given $\rho$, is operationally disturbing to a subsequent measurement of $B$ iff the statistics of $B$ differ for $\rho$ and $\rho '$.
\end{definition}

Moreover, any measure of disturbance should assign the value $0$ to operationally non-disturbing measurements, which is the central \emph{operational requirement} (OR) of this work. This is clearly an uncontroversial demand, however the reason we spell it out explicitly here is precisely that there are recent prominent examples \cite{ozawa2003universally,erhart2012experimental,baek2013experimental,sulyok2013violation} in the literature that fail to adhere to this basic requirement. In this paper we show that satisfying the OR, within quantum theory, rules out a broad class of ``natural'' error-disturbance relations. To show this, we shall prove that for any finite dimensional quantum system, and any two non-commuting observables $A$ and $B$ there always exist pure states $\{\ket{\psi_i}\}$, such that a perfect (projective and sharp) measurement of $A$ can be performed (so there is no error in the statistics of $A$) and the disturbance (in the subsequent statistics of $B$) vanishes. Moreover, we show that the expectation value of the commutator $[A,B]$ for such a state $|\psi_i\rangle$ generically does not vanish.

These results have strong implications for state-dependent formulations of the error-disturbance trade-off relation. To see this, recall that the original Heisenberg argument suggests that $\epsilon(X)\eta(P)\sim h/2$, where $\epsilon(X)$ denotes the error of the approximate position measurement and $\eta(P)$ is the disturbance to the subsequent measurement of momentum. One might heuristically expect a Robertson-like relation $\epsilon(A)\eta(B) \ge |\bra{\psi}[A,B]\ket{\psi}|/2$ to bound the error and disturbance of sequential measurements of arbitrary operators $A$ and $B$ performed in a given state $\ket{\psi}$ (as it was suggested in \cite{ozawa2003universally}), although it should be emphasized this was never claimed by Heisenberg \cite{heisenberg1949physical}. In general, most state-dependent trade-off relations use the expectation value of the commutator in a given state to bound some function of error and disturbance for that state. However, as mentioned, we prove that for every pair of observables $A$ and $B$ there exist pure states for which $\epsilon(A)$ and $\eta(B)$ both vanish, while the expectation value of commutator $\langle \psi | [A, B] |\psi\rangle$ in that state is nonzero. Therefore, any state-dependent error-disturbance relation for sequential measurements of $A$ and $B$ that uses the expectation value of $[A,B]$ in a given state as a lower bound must violate the operational requirement. In other words, the measures of disturbance used in all such relations must take nonzero values even in the situations, when the measurement statistics have not been changed, which is an unphysical conclusion. We illustrate this explicitly by analyzing the state-dependent error-disturbance relation obtained by Ozawa \cite{ozawa2003universally} that has recently received considerable attention.

The paper is organized as follows. In Sec. \ref{sec:requirements} we clarify some confusing aspects of error and disturbance of sequential measurements and explain our approach in detail. Next, in Sec. \ref{sec:znzd}, we describe three families of states for which the error and disturbance can vanish for a given pair of non-commuting observables and emphasize the consequences of the existence of such states for error-disturbance trade-off relations. Section \ref{sec:ozawa} contains the analysis of Ozawa's uncertainty relation, while Sec. \ref{sec:conclusion} concludes the paper.

\section{General framework and the operational requirement}
\label{sec:requirements}

We begin by clarifying some aspects of the error and disturbance in sequential measurements, that, although already described in the literature \cite{busch2004noise}, may be a source of confusion. Specifically, we discuss the merits of state-dependent notions of error and disturbance over state-independent ones, and differentiate between disturbance of a state and disturbance of a measurement. We also present the operational requirement that all operationally-meaningful notions of state-dependent disturbance should satisfy and give a physical justification for it.

\subsection{State-dependent notions}

A state-dependent approach to error and disturbance is based on the following scenario. One is given an initial quantum state of the system, $\rho$, and asks how much an approximate positive operator-valued measure (POVM) measurement $\mathcal{E}_A$ on this state fails to reproduce the perfect measurement of observable $A$, and how much it disturbs the subsequent measurement of the observable $B$. Hence any state-dependent measures of error $\epsilon$ and disturbance $\eta$ depend on three aspects: the approximate measurement $\mathcal{E}_A$ used, the observable to be measured ($A$ or $B$), and the initial state of the system $\rho$. This setting is very broad, and also gives rise to state-independent notions of error and disturbance, for example through averaging over all possible initial states or by finding the maximum and minimum values of $\epsilon$ and $\eta$ over the full set of states \cite{busch2013heisenberg}. In addition, the prior knowledge of the measured state can be utilized. For example,  given a qubit system in an unknown state, then, on average, the projective measurement $\sigma_z$ will disturb the subsequent measurement of $\sigma_x$. However, if one knows that a qubit system is prepared in one of the two eigenstates of the measured $\sigma_z$ operator, then the subsequent measurement of $\sigma_x$ will not be disturbed, as the first measurement clearly does not change the system state.

The goal of state-dependent trade-off relations between error and disturbance is to put a bound on some function of $\epsilon$ and $\eta$, that holds for all approximate measurements $\mathcal{E}_A$ performed on a system prepared in a given state $\rho$. An example of such relation is the already mentioned Robertson-like modification of the original Heisenberg noise-disturbance uncertainty relation, that was proposed by Ozawa \cite{ozawa2003universally},
\begin{equation}
\label{eq:heis}
\epsilon(A,\rho)\eta(B,\rho)\geq\frac{\left|\tr{\rho[A,B]}\right|}{2}.
\end{equation}
In what follows we will refer to this as the restricted Ozawa relation, as it follows from Ozawa's relation if one considers a subset of measurement interactions that are of ``independent intervention'' for the pair $(A,B)$ \cite{ozawa2003universally}. It posits that any approximate measurement on a state $\rho$ may reproduce the ideal projective measurement of observable $A$ on this state with precision limited by noise $\epsilon$, only if it also produces a disturbance $\eta$ of the subsequent projective measurement of $B$, such that the product of $\epsilon$ and $\eta$ is lower-bounded by the average value of the commutator $[A,B]$ in the considered state $\rho$. Such state-dependent formulation of the trade-off relations between error and disturbance of sequential measurements seems to be a commonly chosen approach and one of the most recent results in this field is the ``universally valid error-disturbance uncertainty relation" derived by Ozawa \cite{ozawa2003universally}, which is addressed in Sec. \ref{sec:ozawa}.

\subsection{Disturbance of a measurement: operational requirement}
\label{subsec:minimal}

As the measurement outcomes themselves cannot be controlled, when one describes the disturbing effect of the measurement $\mathcal{E}_A$ on the the subsequent measurement of $B$, it is reasonable to consider the average over all outcomes. That means that one is considering the effect of a nonselective POVM measurement $\mathcal{E}_A$ -- in this way one captures the disturbing effect of the measurement itself, independently of which outcome is recorded. Furthermore, it is important to make a strict distinction between the disturbance of a quantum state and the disturbance to a subsequent quantum measurement. In general, performing a projective measurement of some observable $A$ on a state $\rho$ will affect (disturb) the state and change it into $\rho '\neq\rho$ (apart from the special case, when $\rho$ is diagonal in the basis of eigenstates of $A$). The same holds true for the POVM measurement $\mathcal{E}_A$. The trade-off between information gain and state disturbance is itself a very subtle subject \cite{fuchs1998information,fuchs1996quantum}, especially from the viewpoint of quantum information processing. 

However, let us emphasize that we are interested in the disturbance to a subsequent measurement of $B$ and not of the system state. As even a perfect (projective and sharp) measurement of observable $B$ gives us only insight into the probability distribution of a state $\rho$ over the eigenstates of $B$, any state disturbance causing solely a change of the relative phases between eigenstates of $B$ (the off-diagonal terms) should not be treated as disturbance to the measurement of $B$. In other words, disturbance of the measurement of $B$ occurs if and only if diagonal elements of $\rho$ in the basis of eigenstates of $B$ change. This is the essence of the OR, which is operationally motivated by the fact that only the change in the measurement statistics can be detected by the measurement (otherwise one would call a measurement disturbed even though it is indistinguishable from the perfect one). To be more precise let us denote the outcome probability distribution of a perfect measurement of $B$ in a state $\rho$ by $p_B(\rho)$ and the outcome probability distribution of a measurement of $B$ on a state $\rho '$, obtained after the projective measurement of $A$ performed on the original state $\rho$, by $\tilde{p}_B(\rho)$. Then our requirement can be written as 
\begin{equation}
p_B(\rho)=\tilde{p}_B(\rho)\Leftrightarrow\eta(B|\mathcal{E}_A,\rho)=0,
\end{equation}
which is simply the mathematical expression of the OR.

The requirement that the Born statistical formula be satisfied for perfect (not disturbed) measurements seems to be commonly accepted \cite{busch2004noise, ozawa2003universally, busch2013proof}. The only issue one may worry about is that usually a perfect measurement is defined in a state-independent manner, i.e., that the Born statistical formula should be satisfied for all initial states. As explained before, however, this is not a problem, as the state-independent results can always be recovered from the state-dependent ones. 

\section{Vanishing error and disturbance}
\label{sec:znzd}

As stated, the definition of state-dependent disturbance should only depend upon operational distinguishability between the outcome statistics of a disturbed measurement and the ideal one. Of course one can define an infinite number of different distance measures between probability distributions, however all of them must assign zero for a pair of identical probability distributions. In this section we will investigate the consequences of this for definitions of disturbance, i.e., we will analyze the possibilities and conditions for the perfect measurement of one observable to be performed that causes vanishing disturbance to the measurement of the other (non-commuting) observable on zero-noise, zero disturbance states (ZNZD). The results presented here are thus general and constrain any state-dependent definitions of disturbance that fulfill the OR.

We focus on the measurements in a finite $d$-dimensional Hilbert space and consider the following sequential measurement scenario. We perform the perfect (projective and sharp) measurement of an observable $A$, 
\begin{equation*}
A=\sum_{n=1}^da_n\ketbra{a_n}{a_n},
\end{equation*}
and look at the disturbance to the measurement of the observable $B$, 
\begin{equation*}
B=\sum_{n=1}^db_n\ketbra{b_n}{b_n},
\end{equation*}
that this projective measurement of $A$ induces. The observables $A$ and $B$ are assumed to have non-degenerate spectra only for the sake of clarity, as this is not a necessary condition for the results presented in this section, unless stated otherwise. 

\subsection{The existence of pure ZNZD states}
\label{subsec:existence}

Before presenting the main result of this section consider two cases that one might consider trivial. First, whenever the system state $\rho$ is diagonal in the basis of eigenstates of the measured observable $A$, the state after the measurement, $\rho '$, is clearly not disturbed,
\begin{equation*}
\rho '=\sum_{n=1}^d \bra{a_n}\rho\ket{a_n}\ketbra{a_n}{a_n}=\rho,
\end{equation*}
so also no subsequent measurement is disturbed. That means that in such a case $\forall B:\eta(B)=0$. The second trivial case is when the system is in a maximally mixed state $\rho=\iden/d$, so that it is diagonal in every basis. Since the maximally mixed state is unchanged by any measurement, and every measurement has the same uniform outcome probability distribution,\footnote{Note, however, that the uncertainty here is entirely classical, and not associated with noncommutativity of $A$ and $B$. See \cite{luo2005quantum,korzekwa2014quantum} for subtleties in splitting uncertainty in classical and quantum parts.} one has $\forall A,B:\eta(B)=0$. We also note that both cases trivially satisfy the restricted Ozawa (and any commutator-based) relation, in the sense that the noise, disturbance, and expectation value of $[A,B]$ all simultaneously vanish for these states. The existence of these trivial ZNZD states is therefore consistent with the commutator-based bounds.

However, we can now ask if there exist ZNZD states for which the average value of the commutator $[A,B]$ does not vanish. If the answer to this question is positive, then it means that state-dependent trade-off relations between error and disturbance satisfying the OR cannot be based only on the expectation value of the commutator $[A,B]$. In what follows we show that the answer is in fact positive by proving that for every pair of noncommuting observables $A$ and $B$ there exists a set of pure ZNZD states, and that the expectation value of $[A,B]$ generically does not vanish on these states.

We start by giving a definition of a ZNZD state:
\begin{definition}
A state $\rho$ is a zero-noise, zero-disturbance (ZNZD) state with respect to observables $A$ and $B$ if the perfect (projective and sharp) measurement of an observable $A$ does not change the probability distribution of a subsequent projective measurement of $B$.
\end{definition}
\noindent We now have the following straightforward lemma:
\begin{ZNZD_existence}
If for any two observables $A$ and $B$ there exists a pure state that is unbiased in both bases of eigenstates of $A$ and $B$, then for any two observables $A$ and $B$ there always exists a ZNZD state, i.e.,
\begin{eqnarray*}
&\left(\forall A,B~\exists\ket{\psi_\star}:\forall n \,\,\, |\braket{a_n}{\psi_\star}|^2=|\braket{b_n}{\psi_\star}|^2=\frac{1}{d}\right)&\\
&\Rightarrow\left(\forall A,B~\exists\ket{\psi_\star}:\ket{\psi_\star}~is~a~ZNZD~state\right).&
\end{eqnarray*}
\end{ZNZD_existence}
\begin{proof}
After the projective measurement of $A$ the system initially in a state \ket{\psi_\star} will be transformed into a maximally mixed state. Therefore the outcome probability distribution of the subsequent measurement of $B$ will be uniform, which is the same as before the measurement of $A$, so the disturbance $\eta(B)$ will vanish.
\end{proof}
\noindent Now, in order to prove that such pure ZNZD states exist, we need to prove that the left hand side of Lemma 1 is true. We can now establish the following result.
\begin{existence_theorem}
For any two bases $\{\ket{a_n}\}$ and $\{\ket{b_n}\}$ of a $d$-dimensional Hilbert space there exist at least $2^{d-1}$ states $\ket{\psi_\star}$ that are unbiased in both bases, i.e.,\small
\begin{equation*}
\forall \{\ket{a_n}\},\{\ket{b_n}\}~\exists\ket{\psi_\star}:\forall n |\braket{a_n}{\psi_\star}|^2=|\braket{b_n}{\psi_\star}|^2=\frac{1}{d}.
\end{equation*} \normalsize
\end{existence_theorem}
\begin{proof}
Let $U^\dagger=\sum_n\ketbra{b_n}{a_n}$ denote the unitary connecting the $\{|a_n\rangle \}$ basis to the $\{|b_n\rangle \}$ basis. It is required to show that there exists a pure quantum state $|\psi_\star \rangle$ such that 
\begin{subequations}
\begin{eqnarray}
|\langle a_n|\psi_\star \rangle |^2 &=& \frac{1}{d}, \\
|\langle a_n |U|\psi_\star \rangle |^2 &=& \frac{1}{d}.
\end{eqnarray}
\end{subequations}
The first condition implies that such a state must take the form $|\psi_\star \rangle = \frac{1}{\sqrt{d}} \sum_n e^{i\phi_n} |a_n \rangle$, while the second implies that $U|\psi_\star \rangle =\frac{1}{\sqrt{d}} \sum_n e^{i\gamma_n} |a_n \rangle$. Now the set of such states $|\psi\rangle$ obeying the first condition define a (Lagrangian) torus in the phase variables $\{\phi_n\}$, and moreover, it can be shown \cite{degosson2006symplectic} that the action of the unitary group induces a Hamiltonian flow on the complex projective space $\mathbb{C}P^{d-1}$. However it is known \cite{cho2004holomorphic, lisi2011given} that this torus, when projected into $\mathbb{C}P^{d-1}$, is not ``Hamiltonian displaceable,'' meaning that the image of the torus resulting from the action of $U$ must intersect the original torus (in at least $2^{d-1}$ points). This immediately implies the existence of at least $2^{d-1}$ pure quantum states $\{|\psi_\star \rangle\}$ that satisfy the required conditions.
\end{proof}
It is also clear that the above result cannot extend unconditionally to non-uniform distributions. Specifically, for any given state $\ket{\psi}=\sum_nc_n\ket{a_n}$, such that not all of $\{|c_n|^2\}$ are equal to $1/d$, there will exist a basis $\{\ket{b_n}\}$ in which the probability distribution will differ from the one given by $\{|c_n|^2\}$. To see this let us consider a qubit system with the outcome probability distribution of $\sigma_z$ measurement $(p,1-p)$. States corresponding to this statistics form a circle on the Bloch sphere. Now it is clear, that if $p\neq 1/2$, i.e., if we are not dealing with the great circle, one can find a rotation of the Bloch sphere, such that its action will transform the considered circle to the one not intersecting with the initial one. However, if we limit to ``small rotations,'' so that the ``distance'' between two bases $\{a_n\}$ and $\{b_n\}$ is $R$ (with respect to some appropriately defined distance measure, e.g., $||\iden-U ||$ in the operator norm, where $U$ is the connecting unitary), then for any distribution $p=(p_1, p_2 ,....p_N)$ with $\min(p) > h(R)$ (for some function $h$) there will indeed exist a state $\ket{\psi_\star}$ such that $\ket{\psi_\star}$ has the same statistics with respect to $\{\ket{a_n}\}$ and $\{\ket{b_n}\}$. We leave the precise formulation of this for arbitrary dimensions as an interesting open question.

\subsection{\texorpdfstring{Examples of non-trivial ZNZD states and the generic non-vanishing of $\langle [A,B]\rangle$}{Examples of non-trivial ZNZD states and the generic non-vanishing of <[A,B]>}}

We are now in the position that we know for any observables $A$ and $B$ for a finite-dimensional system, that a ZNZD state $\ket{\psi_\star}$ exists, but we lack the construction of such a state. Therefore it is not \textit{a priori} obvious whether $c=|\bra{\psi_\star}[A,B]\ket{\psi_\star}|$ is nonzero when $[A,B]\neq 0$. 

One particularly simple example is the special case of complementary (mutually unbiased) observables,we have that the eigenbases are related as
\begin{equation}
\forall n: \ket{a_n}=\frac{1}{\sqrt{d}}\sum_{m=1}^de^{i\phi_{mn}}\ket{b_m}.
\end{equation}
Now it is known that for every $d$-dimensional Hilbert space there exist at least three mutually unbiased bases \cite{durt2010mutually}, which means that apart from $\{\ket{a_n}\}$ and $\{\ket{b_n}\}$ bases there also exists a basis $\{\ket{c_n}\}$, such that any $\ket{c_n}$ can be taken as $\ket{\psi_\star}$. Since the construction of three mutually unbiased bases is known, e.g., by using the Heisenberg-Weyl group method \cite{durt2010mutually}, one can simply check if the expectation of the commutator $c$ is nonzero. In the simplest case of $d=2$, the mutually unbiased observables are $A=\sigma_x$ and $B=\sigma_y$, and $\ket{\psi_\star}$ can be chosen from the third unbiased bases formed by the eigenstates of $\sigma_z$. Since $[\sigma_x,\sigma_y]=2i\sigma_z$, therefore the average value of the commutator does not vanish for $\ket{\psi_\star}$ state and is equal to $c=2$. For $d=3$ one can choose the following three unbiased bases:
\begin{eqnarray*}
\{\ket{a_n}\}&=&\left\{(1,0,0),(0,1,0),(0,0,1)\right\},\\
\{\ket{b_n}\}&=&\left\{\frac{1}{\sqrt{3}}(1,1,1),\frac{1}{\sqrt{3}}(1,\omega_3,\omega_3^2),\frac{1}{\sqrt{3}}(1,\omega_3^2,\omega_3)\right\},\\
\{\ket{c_n}\}&=&\left\{\frac{1}{\sqrt{3}}(1,\omega_3^2,\omega_3^2),\frac{1}{\sqrt{3}}(1,\omega_3,1),\frac{1}{\sqrt{3}}(1,1,\omega_3)\right\},
\end{eqnarray*}
where $\omega_3=\exp(2\pi i/3)$. In this case $c$ also does not vanish for at least one of the $\ket{c_n}$ states, unless $[A,B]=0$, which can only be the case when $A$ or $B$ is completely degenerate and thus proportional to identity. As an example let us choose eigenvalues of $A$ and $B$ to be $a_1=b_1=-1$, $a_2=b_2=0$, $a_3=b_3=1$. Then $c=1/\sqrt{3}$ (for $\ket{\psi_\star}\in\{\ket{c_1},\ket{c_2}\}$) or $c=2/\sqrt{3}$ (for $\ket{\psi_\star}=\ket{c_3}$). Similarly, for \mbox{$d=4$} one can choose the eigenstates of $A$, $\{\ket{a_n}\}$, to be the two-qubit computational basis, and the eigenstates of $B$ to be defined by \mbox{$\{\ket{b_n}=H\otimes H \ket{a_n}\}$}, where $H$ is the two-dimensional Hadamard matrix. These two bases are mutually unbiased and, since for \mbox{$d=4$} there exist five mutually unbiased bases, it leaves 12 states (four from each of the remaining three bases) that are ZNZD states with respect to $A$ and $B$. Again, unless $[A,B]=0$, at least for one of these states the expectation value $c$ of the commutator does not vanish.

Beyond the low-dimensional examples presented, it is clear that $\bra{\psi_\star}[A,B]\ket{\psi_\star}$ does not vanish in general unless we make a special choice of eigenvalues, for example by making some of them degenerate. However, being given eigenstates of two observables and the freedom to choose their eigenvalues, one can always make $c$ non-vanishing for unbiased states $\ket{\psi_\star}$. Indeed, it is clear to see that $\bra{\psi_\star}[A,B]\ket{\psi_\star}=0$ corresponds to a set of measure zero in the space of eigenvalues. 

\subsection{Consequences for noise-disturbance relations}
\label{subsec:consequences}

As already mentioned, the existence of pure ZNZD states $\ket{\psi_\star}$ for every pair of non-commuting observables $A$ and $B$, such that the average of $[A,B]$ does not vanish, implies that any relation of the form
\begin{equation}
\sum_{m,n=0}^\infty f_{mn}(A,B) \epsilon^m(A,\rho)\eta^n(B,\rho)\geq \left|\tr{\rho[A,B]}\right|,
\end{equation} 
with $f_{00}=0$, must violate the OR. This includes the restricted Ozawa relation, Eq. (\ref{eq:heis}), as well as the Ozawa's ``universally valid error-disturbance uncertainty relation'' given by 
\begin{equation}
\label{eq:ozawa}
\epsilon(A)\eta(B)+\epsilon(A)\sigma(B)+\sigma(A)\eta(B)\geq\frac{\left|\tr{\rho[A,B]}\right|}{2},
\end{equation}
where the dependence on $\rho$ of all of the terms on the left-hand side was omitted to shorten the notation and $\sigma(A)$ denotes the standard deviation of the outcome statistics of $A$.

\section{Analysis of the Ozawa error-disturbance relation}
\label{sec:ozawa}

The aim of this section is to show that the well-known state-dependent Ozawa's trade-off relation \cite{ozawa2003universally}, given by Eq. (\ref{eq:ozawa}), violates the OR, and so care should be taken in its interpretation. We decided to discuss this particular relation separately from the general case presented in the previous section, due to the recent experimental investigations of Ozawa's error-disturbance trade-off relation with the use of qubit systems \cite{erhart2012experimental,baek2013experimental,sulyok2013violation}. These works claim to experimentally verify the Ozawa's relation, which implies that any measurement of an observable $A$ in a state $|\psi\rangle$ with error $\epsilon(A)$ causes disturbance $\eta(B)$ on another observable $B$ satisfying Eq. (\ref{eq:ozawa}). What we want to emphasize here is that the notion of disturbance being used has the non-operationally motivated properties, and so the sense in which it can describe how a disturbed measurement on any given state differs from the perfect one is debatable. If one insists on using what we consider better operationally motivated definition, then such a trade-off between error and disturbance will not be inevitable in general (i.e. applying for all states). 

To see this more clearly let us analyze the Ozawa relation, specifically the experimentally investigated case of $d=2$. Central to the relation are the error and disturbance terms which can be defined respectively for two observables $A$ and $B$ and a pure state $|\psi\rangle$ as \cite{ozawa2005universal}
\begin{subequations}
\begin{eqnarray}
\epsilon_{\mathrm{O}}(A)^2 &=& \sum_k || M_k (m -A) |\psi\rangle ||^2,\label{ozawadefs1}\\
\eta_{\mathrm{O}}(B)^2 &=& \sum_k || [M_k ,B] |\psi\rangle ||^2,\label{ozawadefs2}
\end{eqnarray}
\end{subequations}
where $\{M_k\}$ are the POVM elements induced by the actual measurement performed on the system, and $m$ denote the corresponding eigenvalues of the observable. These terms, together with the variances $\sigma(A) $ and $\sigma(B)$ of $A$ and $B$ in the state $|\psi\rangle$ can be shown to obey the error-disturbance relation given by Eq. (\ref{eq:ozawa}), which is argued to be a rigorous generalization of Heisenberg's microscope relation \cite{ozawa2003universally}. The above measures of error and disturbance, given in Eqs. (\ref{ozawadefs1}) and (\ref{ozawadefs2}), have been accused of being problematic, both in terms of what they quantify \cite{busch2013proof, busch2004noise} and in relation to interpretative issues \cite{dressel2013certainty} (see \cite{busch2013noise} for a recent and extensive critique). Here we address the (non)operational meaning of the disturbance term $\eta_{\mathrm{O}}$ as well as its apparent state dependence.

First of all, let us note that if a perfect (projective and sharp) measurement of observable $A$ is performed on a state $\ket{\psi}$ then \mbox{$\epsilon_{\mathrm{O}}(A)=0$} and
\begin{equation*}
\eta_{\mathrm{O}}(B)^2 = \sum_k || [\ket{a_k}\!\bra{a_k},B] |\psi\rangle ||^2.
\end{equation*}
Focusing on the disturbance for the initial state of the system being $\ket{a_l}$, i.e., the eigenstate of $A$, one has
\begin{equation*}
\eta_{\mathrm{O}}(B)^2 = \sum_{k\neq l} || \ket{a_k}\!\bra{a_k} B\ket{a_l} ||^2+|| (\bra{a_l}B\ket{a_l}-B)\ket{a_l} ||^2.
\end{equation*}
The sum on the right-hand side of the above equation vanishes only when $\ket{a_l}$ is the eigenstate of $B$ (as $B\ket{a_l}$ must be orthogonal to all $\ket{a_k}$). Therefore, unless all the eigenstates of $A$ coincide with the eigenstates of $B$ (which implies \mbox{$[A,B]=0$}), for at least one of such eigenstates the disturbance is nonzero. We identify this as a very unphysical property of the disturbance measure, as the measurement of $A$ performed on the eigenstate of $A$ not only does not change the outcome probability distribution of the subsequent measurement of $B$, but also does not change the state of the system at all. 

Now let us turn to the qubit scenario. It is easy to compute \cite{erhart2012experimental} that $\eta_{\mathrm{O}}$ for the sequential projective measurements of \mbox{$A=\v{a}\cdot\v{\sigma}$} and \mbox{$B=\v{b}\cdot\v{\sigma}$} (with $\v{\sigma}$ denoting the vector of Pauli matrices) on a qubit system in state $\ket{\psi}$ is given by
\begin{eqnarray}
\eta_{\mathrm{O}}(B,\ket{\psi})&=&\sqrt{2}\left|\sin\beta\right|,
\end{eqnarray}
where $\beta$ is the angle between the Bloch vectors $\v{a}$ and $\v{b}$. For this primitive scenario, we find that although the definition of $\eta_{\mathrm{O}}$ appears to be state dependent, the resultant expression for a qubit system turns out to have no dependence on the system state $\ket{\psi}$. Further insight can be obtained by the following observation. Let us introduce the state-dependent measure of disturbance $\eta_{\mathrm{K}}$ defined by the Kolmogorov distance between outcome probability distributions of a perfect and disturbed measurement, i.e.,
\begin{equation}
\eta_{\mathrm{K}}(B,\ket{\psi}):=K(p_B,\tilde{p}_B)=\frac{1}{2}\sum_n|p^{(n)}_B-\tilde{p}^{(n)}_B|,
\end{equation}
where the dependence of $p_B$ and $\tilde{p}_B$ on $\ket{\psi}$ was omitted to shorten the notation. The operational meaning of the introduced measure of disturbance is as follows: the optimal success probability with maximum likelihood estimation for distinguishing between the perfect and disturbed probability distributions is given by \mbox{$[1+K(p_n,\tilde{p}_n)]/2$}. Now it can be shown that the expression for Ozawa's disturbance $\eta_{\mathrm{O}}$ can be recovered by averaging the disturbance $\eta_K$, over all possible states of the system,
\begin{equation*}
\langle\eta_{K}(B,\ket{\psi})\rangle_{\rm Bloch}=\frac{1}{4}\left|\sin\beta\right|=\frac{1}{4\sqrt{2}}\eta_{\mathrm{O}}(B,\ket{\psi}).
\end{equation*}
Thus for $d=2$ the definition of disturbance proposed by Ozawa coincides with the average over the state-dependent notion defined here. It follows that $\eta_{\mathrm{O}}(B,\ket{\psi})$ does not satisfy the operational requirement. 

Finally, let us note that in a qubit case the set of non-trivial ZNZD states is not only limited to states $\ket{\psi_{\star}}$, i.e., the states unbiased in the bases of eigenstates of $A$ and $B$. Without the loss of generality one may choose the Bloch vectors representing considered observables to be \mbox{$\v{a}=(0,0,1)$} and \mbox{$\v{b}=(\sin\beta,0,\cos\beta)$}. Then one can easily show that a projective measurement of $A$ on any of the states represented by the Bloch vector \mbox{$\v{c}=(0,\sin\theta,\cos\theta)$} does not change the statistics of the subsequent measurement of $B$. Therefore the disturbance $\eta(B)$ caused by the projective measurement of $A$ for all such states should vanish.

\section{Outlook}
\label{sec:conclusion}

In this paper we have tried to highlight some subtleties of sequential measurements in finite-dimensional Hilbert spaces by examining state-dependent notions of disturbance. A core element of our reasoning relies on the insistence on basic operational principles, in particular the operational requirement (OR), which states that a measurement cannot be treated as disturbed if its outcome statistics is identical to the one for the perfect measurement. By defining zero-noise, zero-disturbance (ZNZD) states and proving the existence of such pure states with nonvanishing expectation value of the commutator of measured observables, we have shown that no traditional commutator-based bound for the state-dependent trade-off between error and disturbance can hold for all states, while also satisfying the OR. We have also addressed one of the recent formulations of the error-disturbance uncertainty relation derived by Ozawa, pointed out the unphysical  properties of disturbance used in this approach and shown that in the single qubit case Ozawa's disturbance can be obtained via uniform averagings of state-dependent disturbance over the Bloch sphere.

One may be tempted to introduce an operationally motivated requirement also for the error of measurement, similarly to the OR. Let us however note that, due to the state-dependent nature of considered relations, this leads to problematic issues. To see this, consider the following requirement: \emph{a measurement $\mathcal{E}_A$ that perfectly reproduces the measurement statistics of observable $A$ in a given state $\rho$ should not be called noisy}. If there are no restrictions on the measurements and given a state $\rho$ one can always choose a ``simulating measurement'' with POVM elements $M_n=\sqrt{p_n}\iden$ for $n=1\dots d$ and $p_n=\tr{\rho\ketbra{a_n}{a_n}}$. Such a measurement does not affect the system state, so it does not disturb subsequent statistics of any measurement, and it also perfectly reproduces the measurement of $A$ for the state $\rho$. The existence of such measurement clearly shows that it is not only error and disturbance, but also the information-gain about the system, that must be considered in a state-dependent trade-off relation (in the above example both error and disturbance vanish, but there is also no information-gain). 

The origin of complementarity and the error-disturbance trade-off lies in the noncommutativity of the measured observables. Our main result, however, states that there cannot exist a simple state-dependent relation connecting the trade-off between error and disturbance with the expectation value of the commutator in the considered state. A more tractable line to follow is to relate error, disturbance and non-commutativity of the measured observables in a Heisenberg-Robertson-like inequality, in which both the error and disturbance measures are state-independent quantities. A recent example of such an approach to sequential measurements in finite-dimensional Hilbert spaces, specifically for single qubit observables, was recently presented in Ref. \cite{busch2013heisenberg}.

\textbf{Acknowledgements:} We would like to thank C. Branciard for his useful comments on non-informative POVMs, as well as P. Busch, P. Lahti and R. F. Werner for their helpful comments on our manuscript. This work was supported by EPSRC. D.J. was supported by the Royal Society. T.R. was supported by the Leverhulme Trust.

\bibliographystyle{prsty}
\bibliography{bibliography}

\begin{thebibliography}{10}

\bibitem{robertson1929uncertainty}
H.~P. Robertson, Phys. Rev. {\bf 34},  163  (1929).

\bibitem{heisenberg1949physical}
W. Heisenberg, {\em The Physical Principles of the Quantum Theory} (Courier
  Dover Publications, Mineola, 1949).

\bibitem{ozawa2003universally}
M. Ozawa, Phys. Rev. A {\bf 67},  042105  (2003).

\bibitem{busch2013proof}
P. Busch, P. Lahti, and R.~F. Werner, Phys. Rev. Lett. {\bf 111},  160405
  (2013).

\bibitem{ozawa2004uncertainty}
M. Ozawa, Phys. Lett. A {\bf 320},  367  (2004).

\bibitem{hall2004prior}
M.~J.~W. Hall, Phys. Rev. A {\bf 69},  052113  (2004).

\bibitem{rozema2013note}
L.~A. Rozema, D.~H. Mahler, A. Hayat, and A.~M. Steinberg, arXiv:1307.3604
  (2013).

\bibitem{ozawa2013disproving}
M. Ozawa, arXiv:1308.3540  (2013).

\bibitem{dressel2013certainty}
J. Dressel and F. Nori, Phys. Rev. A {\bf 89},  022106  (2014).

\bibitem{busch2013heisenberg}
P. Busch, P. Lahti, and R.~F. Werner, Phys. Rev. A {\bf 89},  012129  (2014).

\bibitem{bastos2014robertson}
C. Bastos, A.~E. Bernardini, O. Bertolami, N. Costa~Dias, and J.~a.~N. Prata,
  Phys. Rev. A {\bf 89},  042112  (2014).

\bibitem{ipsen2013error}
A.~C. Ipsen, arXiv:1311.0259  (2013).

\bibitem{busch2013noise}
P. Busch, P. Lahti, and R.~F. Werner, arXiv:1312.4393  (2013).

\bibitem{branciard2013error}
C. Branciard, Proc. Natl. Acad. Sci. USA {\bf 110},  6742  (2013).

\bibitem{ringbauer2013joint}
M. Ringbauer, D.~N. Biggerstaff, M.~A. Broome, A. Fedrizzi, C. Branciard, and
  A.~G. White, Phys. Rev. Lett. {\bf 112},  020401  (2014).

\bibitem{busch2004noise}
P. Busch, T. Heinonen, and P. Lahti, Phys. Lett. A {\bf 320},  261  (2004).

\bibitem{erhart2012experimental}
J. Erhart, S. Sponar, G. Sulyok, G. Badurek, M. Ozawa, and Y. Hasegawa, Nat.
  Phys. {\bf 8},  185  (2012).

\bibitem{baek2013experimental}
S.-Y. Baek, F. Kaneda, M. Ozawa, and K. Edamatsu, Sci. Rep. {\bf 3},  2221
  (2013).

\bibitem{sulyok2013violation}
G. Sulyok, S. Sponar, J. Erhart, G. Badurek, M. Ozawa, and Y. Hasegawa, Phys.
  Rev. A {\bf 88},  022110  (2013).

\bibitem{fuchs1998information}
C.~A. Fuchs, Fortschritte der Physik {\bf 46},  535  (1998).

\bibitem{fuchs1996quantum}
C.~A. Fuchs and A. Peres, Phys. Rev. A {\bf 53},  2038  (1996).

\bibitem{luo2005quantum}
S. Luo, Theor. Math. Phys. {\bf 143},  681  (2005).

\bibitem{korzekwa2014quantum}
K. Korzekwa, M. Lostaglio, D. Jennings, and T. Rudolph, Phys. Rev. A {\bf 89},
  042122  (2014).

\bibitem{degosson2006symplectic}
M. De~Gosson, {\em Symplectic geometry and quantum mechanics} (Birkh\"auser
  Verlag, Basel, 2006).

\bibitem{cho2004holomorphic}
C.-H. Cho, Int. Math. Res. Not. {\bf 2004},  1803  (2004).

\bibitem{lisi2011given}
S. Lisi, \emph{Given two basis sets for a finite Hilbert space, does an
  unbiased vector exist?}, http://math.stackexchange.com/q/29819 (2011).

\bibitem{durt2010mutually}
T. Durt, B.-G. Englert, I. Bengtsson, and K. {\.Z}yczkowski, Int. J. Quantum
  Inf. {\bf 8},  535  (2010).

\bibitem{ozawa2005universal}
M. Ozawa, J. Opt. B: Quantum Semiclass. Opt. {\bf 7},  S672  (2005).

\end{thebibliography}

\end{document}